\documentclass[prb,a4paper,showpacs,twocolumn,superscriptaddress,longbibliography]{revtex4-1}

\usepackage{textcomp}
\usepackage{amssymb}
\usepackage{amsmath}
\usepackage{amsfonts}
\usepackage{graphicx}
\usepackage{bm}
\usepackage{xcolor}
\usepackage{multirow}
\usepackage{natbib}
\usepackage{hyperref}
\usepackage{mathrsfs}

\begin{document}

\title{Spintronics with a Weyl point in superconducting nanostructures}

\author{Y. Chen}

\affiliation{Kavli Institute of Nanoscience, Delft University of Technology, 2628 CJ Delft, The Netherlands}

\author{Y. V. Nazarov}

\affiliation{Kavli Institute of Nanoscience, Delft University of Technology, 2628 CJ Delft, The Netherlands}

\begin{abstract}
We investigate transport in a superconducting nanostructure housing a Weyl point in the spectrum of Andreev bound states. A minimum magnet state is realized in the vicinity of the point. One or more normal-metal leads are tunnel-coupled to the nanostructure. We have shown that this minimum magnetic setup is suitable for realization of all common goals of spintronics: detection of a magnetic state, conversion of electric currents into spin currents, potentially reaching the absolute limit of one spin per charge transferred, detection of spin accumulation in the leads. 
The peculiarity and possible advantage of the setup is the ability to switch between magnetic and non-magnetic state by tiny changes of the control parameters: superconducting phase differences. We employ this property to demonstrate the feasibility of less common spintronic effects: spin on demand and alternative spin current. 
\end{abstract}

\maketitle

\section{Introduction}
\label{sec:intro}
Spin currents in metals are conserved at significant length scale of spin-flip length and therefore can be induced and measured at this scale. The whole field of spintronics \cite{SpintronicsReview,BauerFirst} concentrates on conversion of electric currents to spin currents, electrical detection of spin polarization produced by the spin currents, and dynamics of these processes \cite{BauerTwo}. Much theoretical research addressed this conversion and detection at ferromagnet-normal metal interfaces, for collinear\cite{Fert,Silsbee} and non-collinear\cite{BrataasNazarov, BrataasBauer} configurations of the ferromagnets.  Detection of  the complex counting statistics of spin currents has been addressed as well\cite{countingstatistics, countingstatistics2}. 

New functionalities can be achieved by combining ferromagnets, normal metal and superconductors, most are based on spin-singlet nature of Cooper pairs forming the superconducting condensate\cite{Linder2015}. For instance, the absolute spin-valve effect \cite{absolutespinvalve,absoluteexp} can be achieved in this way, and long-distance triplet proximity effect \cite{Longrange,Oddfrequency,Braude} can be arranged. 

While most research and applications in spintronics concentrates on extended structures, all spintronic effects can be reproduced with the systems involving few quantum states, for instance, realized in semiconducting quantum dots \cite{Ludwig2013,Hanson}. Spin filtering and detection have been demonstrated\cite{Hanson,Hanson2,Ono1313} and more research is underway\cite{Bordoloi2020}. The ferromagnets are not needed here since spin effects arise from Zeeman splitting of the discrete energy levels by external magnetic field. 

Recently, Weyl points - the topological singularities in the spectrum of Andreev bound states  - have been predicted in superconducting nanostructures\cite{Weyl}. At a Weyl point, the energy of the lowest Andreev state crosses Fermi level, so it costs vanishing energy to excite a quasiparticle near the Weyl point. From general topological reasoning, such crossing requires tuning of three parameters. This is why the Weyl points are usually considered in multi-terminal superconducting nanostructures where the parameters are the superconducting phase differences of the terminals. Four terminals are thus needed to realize a Weyl point. This prediction gave rise to related experimental and theoretical research \cite{Manucharyan2020, Belzig2020, Pribiag2020, Marra2019, Scherubl2019, Houzet2019, Repin2019, Sonic2019, Finkelstein2019, WeylDisks, Meyer2017, Eriksson2017}.

It is important that weak spin-orbit interaction splits the energies of single-quasiparticle states.\cite{Weyl, Yokoyama} Owing to this, the ground state configuration is always magnetic in a small finite region around the point and is non-magnetic otherwise.\cite{Yokoyama, Repin2019} The opposite magnetization is realized in a small region at opposite settings of the phase differences, as required by time reversibility. Thus Weyl point provides a minimum magnet that involves a single electron spin and can be driven to a non-magnetic state by a tiny change of the external parameters --- superconducting phases. More details are provided in Section \ref{sec:magn}.
\begin{figure}
\begin{center}
\includegraphics[width=\columnwidth]{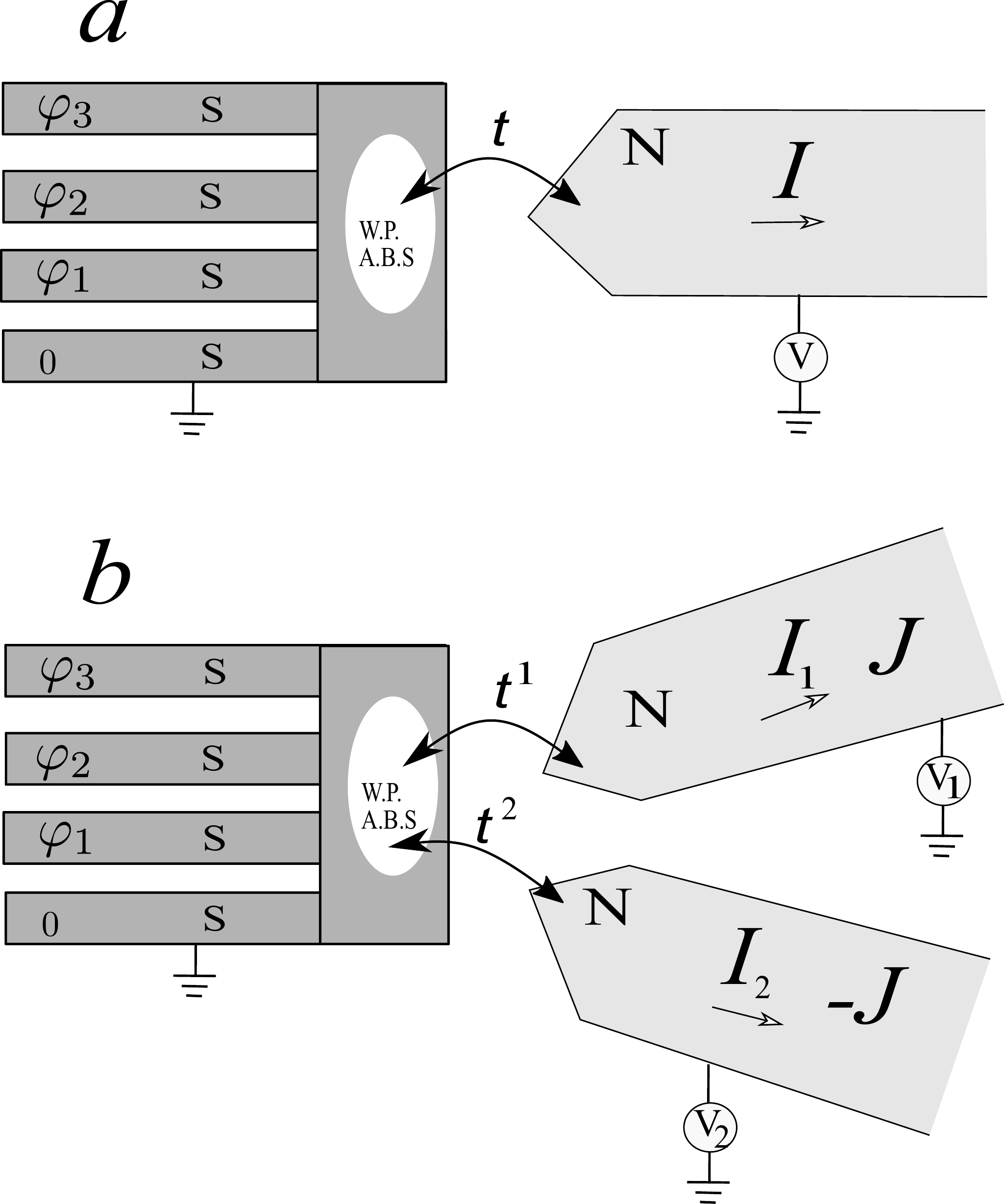}
\end{center}
\caption{\label{fig:setup}Single-lead (a) and two-lead (b) setups for spintronics with a Weyl point investigated in the Article. The normal leads are tunnel-coupled to a superconducting structure hosting an Andreev bound state (W.P.A.B.S. in the Figure) that can be tuned to a Weyl point by choosing the superconducting phases $\phi_{1,2,3}$. We demonstrate the spintronic effects in the transport: spin current $J$ in addition to electric current $I$, and the detection of possible spin accumulation in the leads.}
\end{figure}
 
In this Article, we investigate if this minimum magnet can be utilized in spintronic context. We consider low-voltage transport in a setup where one or two normal leads are tunnel-coupled to a superconducting structure hosting a Weyl point (Fig. \ref{fig:setup}). We demonstrate that this suffice to realize all spintronic effects: the magnetic state of the superconducting structure can be detected, a spin-polarized current can be induced in the leads, and its polarization can be close to absolute one, non-equilibrium spin accumulation in the leads can be detected electrically. The peculiarity and a possible advantage of the Weyl-point spintronics is the sensitivity of all effects to tiny variations of the superconducting phases. This enables spintronic effects that are not usually present in common situations: we discuss how to provide spin on demand and alternative spin current.

The structure of the Article is as follows.
In Section \ref{sec:magn} we review the generic Hamiltonian of the Weyl point and explain the magnetism in its vicinity. In Section \ref{sec:micro} we establish a microscopic model of tunneling to/from the nanostructure, identify the elementary transport processes, compute their rates and derive a master equation describing the transport. We study the transport in a single-lead setup in Section \ref{sec:sing}. Next, we describe how to achieve spin on demand and alternative spin current (Section \ref{sec:spino}). Owing to spin conservation in the superconductor, the d.c. spin current requires two leads: we consider this situation in Section \ref{sec:twol} and show how to approach the absolute spin polarization of the resulting current. We discuss the detection of spin accumulation in the leads in Section \ref{sec:dete}. We conclude in Section \ref{sec:conc}.

\section{Magnetism near a Weyl point}
\label{sec:magn}
In this Section, we will give the effective Hamiltonian of the superconducting nanostructure in the vicinity of Weyl point and describe its magnetic state following the references\cite{Weyl, Yokoyama, Repin2019}. 
\begin{figure}
\begin{center}
\includegraphics[width=\columnwidth]{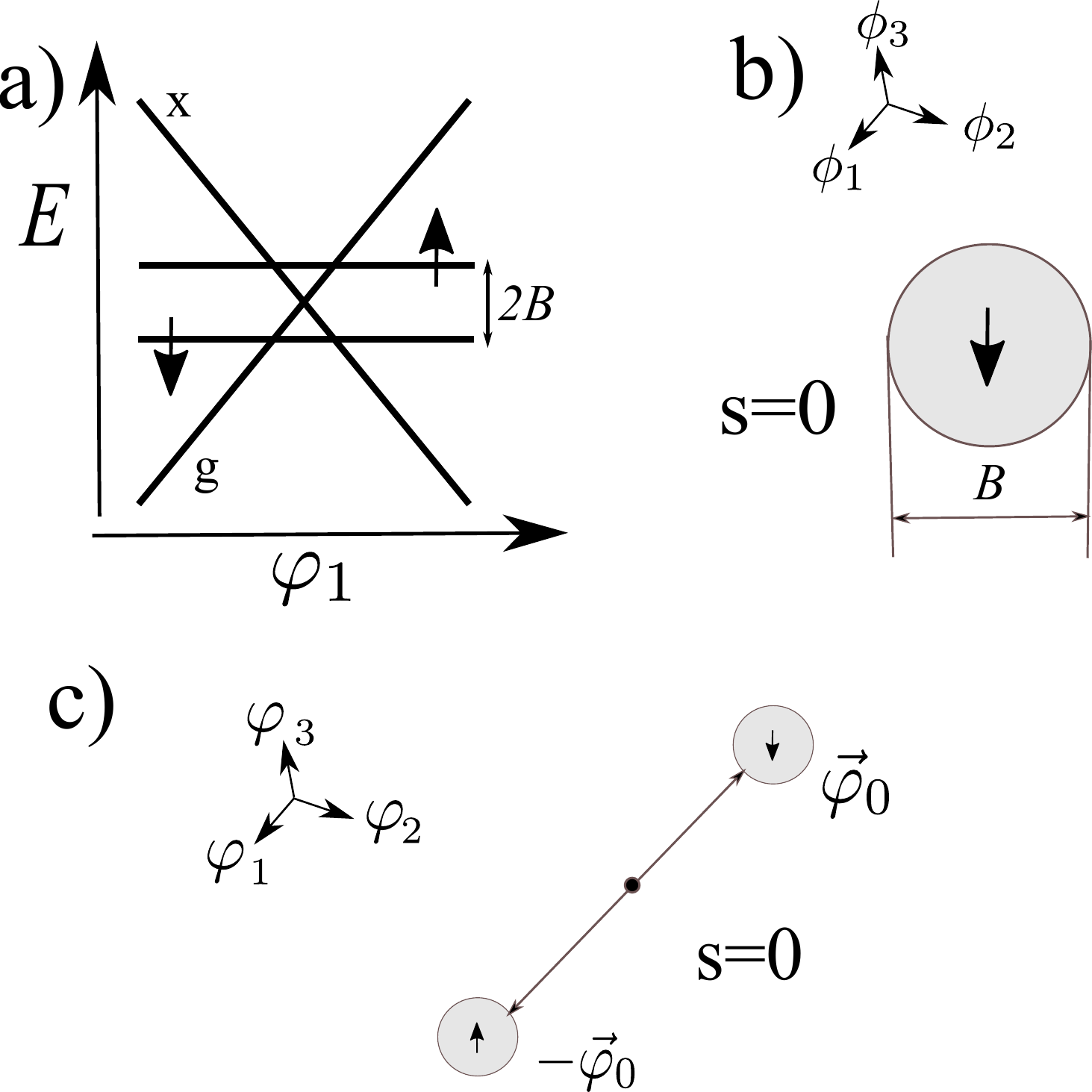}
\end{center}
\caption{\label{fig:magnet}Magnetism near a Weyl point (a) Energies of the singlet ($g$,$x$) and doublet ($\downarrow$, $\uparrow$) states  in the vicinity of a Weyl point versus one of the phases. (b) The region of a doublet (magnetic) ground state around a Weyl point at $\vec{\varphi}_0$ surrounded by the region of a singlet ground state. (c) Opposite magnetizations at Weyl points at $\pm \vec{\varphi}_0$. }
\end{figure}
Three independent superconducting phase differences can be regarded as a 3D vector $\vec{\varphi}$. Suppose the Weyl points are situated at $\pm \vec{\varphi}_0$.  In the vicinity of the point at $\vec{\varphi}_0$ we expand $\vec{\varphi} = \vec{\varphi}_0 + \delta\vec{\varphi}$, $|\delta\vec{\varphi}|\ll 1$ and can describe the lowest Andreev bound states by a $2\times 2$ matrix BdG Hamiltonian
\begin{equation}
\hat{H}_{{\rm W}} = \phi_a \hat{\tau}_a; \; \phi_a = M_{ab} \delta \varphi_b,
\end{equation} 
where $\hat{\tau}_a$ is a vector of Pauli matrices. This form suggests convenient coordinates $\vec{\phi}$ for the vicinity of a Weyl point that are linearly related and thus equivalent to $\delta \vec{\varphi}$. We will make use of these coordinates through the paper. In these coordinates of dimension energy, the spectrum is isotropic and conical, $E = \pm |\vec{\phi}|$. The coordinates are thus defined upon an orthogonal transformation.

The $2 \times 2$ BdG Hamiltonian is obtained by projection on two two-component eigenfunctions $|\Psi_{\pm}\rangle$ related by BdG symmetry. In coordinate representation, 
\begin{equation}
|\Psi_+\rangle = (u({\bf r}),v({\bf r})); \;  |\Psi_-\rangle = (-v^*({\bf r}),u^*({\bf r})),
\end{equation}
${\bf r}$ being the coordinates within the nanostructure.

Weak spin-orbit interaction within the nanostructure modifies the Hamiltonian splitting the Andreev states in spin\cite{Yokoyama},
\begin{equation}
\hat{H}_{{\rm W}} = \phi_a \hat{\tau}_a + B_a \hat{\sigma}_a, 
\end{equation}
$\hat{\sigma}_a$ being a vector of Pauli matrices in spin space, and $B_a$ looks like an external magnetic field causing Zeeman splitting. However, $\vec{B} \ne 0$ even in the absence of external magnetic field and represents the effect of the superconducting phase differences on spin orientation. Owing to global time reversibility, the vectors $\vec{B}$ are opposite for opposite Weyl points, $\vec{B}(-\varphi_0)= - \vec{B}(\varphi_0)$. The magnitude of $\vec{B}$ can be estimated as the superconducting energy gap $\Delta$ times a dimensionless factor characterizing the weakness of the spin-orbit interaction. For a concrete number in mind, we can take $B \simeq 0.1 \Delta \simeq 0.2 meV$ which corresponds to niobium. If there is an external magnetic field, it adds to $\vec{B}$. We note however that our estimation of $B$ is about $3 T$, so it requires a significant field to change it. 

To rewrite the Hamiltonian in the second-quantization form, we introduce quasiparticle annihilation operators $\hat{\gamma}_\sigma$ and associated Nambu bispinors $\bar{\gamma}_{a,\sigma} \equiv (\hat{\gamma}_\sigma, \sigma \hat{\gamma}_{-\sigma}^\dag)$ to recast it to the standard form,
\begin{equation}
H_{{\rm WP}} = \frac{1}{2} \bar{\gamma}^\dagger_\alpha \hat{H}^{{\rm WP}}_{\alpha \beta}  \bar{\gamma}_\beta
\end{equation}
This Hamiltonian can be reduced to a diagonal form for a certain direction in $\phi$-space, $\vec{\phi} = \phi {\vec {n}}$ by a Bogoliubov transform of $\hat{\gamma}_\sigma$ to a direction-dependent $\hat{\tilde{\gamma}}_\sigma$. Choosing the spin quantization axis along $\vec{B}$, we arrive at
\begin{equation}
H_{{\rm WP}} = \frac{1}{2}(\phi +B\sigma) \left(\hat{\tilde{\gamma}}_\sigma^\dagger \hat{\tilde{\gamma}}_\sigma - \hat{\tilde{\gamma}}_\sigma \hat{\tilde{\gamma}}^\dagger_\sigma\right)
\end{equation}
This gives the spectrum sketched in Fig. \ref{fig:magnet}a. The energies are $E = \pm \phi$ for two spin-singlet states, ground one $|g\rangle$, and excited one $|x\rangle \equiv \hat{\tilde{\gamma}}^\dagger_{\uparrow} \hat{\tilde{\gamma}}^\dagger_{\downarrow}|g\rangle$. The energies are  $E = \pm B$ for two components of the spin doublet $|\uparrow\rangle \equiv \hat{\tilde{\gamma}}^\dagger_{\uparrow}|g\rangle$, $|\downarrow\rangle  \equiv \hat{\tilde{\gamma}}^\dagger_{\downarrow}|g\rangle$. The spin-doublet is split and its energies exhibit no singularity or phase dependence in the vicinity of the Weyl point $\vec{\phi}=0$, while the spin-singlet states retain the conical spectrum. 

This leads us to a simple but important conclusion: the ground state of the nanostructure is magnetic in a narrow vicinity of a Weyl point, namely, at $|\phi|<B$ (Fig. \ref{fig:magnet}a). Corresponding to our estimation of $B$, $\delta \varphi \simeq 0.1$. Thus, the magnetism can be switched on and off by variation of magnetic flux controlling the superconducting phase differences by a tenth of the flux quantum. This is a much smaller action than, for instance, in quantum dots where it requires a change of electron number and strong magnetic fields, not mentioning the bulk magnetic structures. The opposite direction of the equilibrium magnetic polarization is found at the opposite Weyl point (Fig. \ref{fig:magnet}c).

This makes a nanostructure with Weyl points a minimum example of a magnet.

\section{Microscopic model and tunneling rates}
\label{sec:micro}
Let us consider tunneling between the electron states in the nanostructure and those in a normal lead. Conventionally, we assume a quasi-continuous spectrum in the lead and label the electron states with $k$ and spin direction $\sigma$, $\hat{d}_{\sigma,k}$ being an associated electron creation operator. We start with a rather general model tunneling Hamiltonian
\begin{equation}
H_T = \int d{\bf r} (t_k({\bf r})\hat{c}_\sigma({\bf r})^\dag\hat{d}_{\sigma,k} +h.c.)
\end{equation}
that describes electron tunnelling to/from a point ${\bf r}$ in the nanostructure from/to the state $k$ in the lead, $\hat{c}_\sigma({\bf r})$ being the electron annihilation operator    at the point ${\bf r}$. We assume spin conservation in the course of tunneling, this is consistent with the assumption of weak spin-orbit interaction. 

To proceed, one represents $\hat{c}_\sigma({\bf r})$ in terms of the quasiparticle creation/annihilation operators $\hat{\gamma}_{\sigma,n}$ associated with the quasiparticle states in the nanostructure, those are labelled with $n$:
\begin{equation}
\hat{c}_\sigma({\bf r}) = \sum_n \left(u_n({\bf r})\hat{\gamma}_{\sigma,n} - \sigma v_n^*({\bf r})\hat{\gamma}^\dag_{-\sigma,n}\right).
\end{equation}  
Here, $(u_n({\bf r}), v_n({\bf r})$ is the wave function of the quasiparticle state $n$.

We concentrate on the tunneling that involves only the lowest quasiparticle state near the Weyl point, this is relevant at low energies $\ll \Delta$. We also neglect higher-order tunneling processes corresponding to two-electron tunneling to the superconducting nanostructure \cite{Frank} or Andreev reflection from the nanostructure. With this, we can replace 
\begin{equation}
c_\sigma({\bf r}) \to  u({\bf r})\hat{\tilde{\gamma}}_\sigma - \sigma v^*({\bf r})\hat{\tilde{\gamma}}^\dag_{-\sigma}
\end{equation}
where $\hat{\tilde{\gamma}}_\sigma$ is the direction-dependent quasiparticle creation operator, and $(u({\bf r}), v({\bf r}))$ is the associated wave function which also depends on the direction ${\bf n}$.

With this, we can express all the tunneling rates involving electron energy $E$ in terms of two combinations of the tunneling amplitudes:
\begin{eqnarray}
\Gamma_{u,v} &=& \frac{2\pi}{\hbar} \sum_{k} \delta(E - E_k) |T^{u,v}_k|^2
 \\
 T^u_k &=& \int d{\bf r}  u({\bf r}) t^*_k({\bf r}) ; \;
 T^v_k = \int d{\bf r}  v({\bf r})  t_k({\bf r})
\end{eqnarray}
Here, $\Gamma_u$ enters the rates of the processes where adding/extracting of an electron in the lead is accompanied by extracting/adding a quasiparticle, while $\Gamma_v$ determines the rates of the processes where the adding/extracting of an electron goes together with the adding/extracting a quasiparticle. These rates depend on the direction in the vicinity of the Weyl point. Transforming the wave functions, we derive the $\vec{n}$ dependence of these rates:
\begin{equation}
\label{eq:angular}
\Gamma_{u,v} = \frac{\Gamma}{2} \pm \vec{\Gamma}_1 \cdot \vec{n}; \; |\vec{\Gamma}_1| < \Gamma/2
\end{equation}
We observe that the tunneling breaks isotropy near the Weyl point. This has been also noted in $\cite{Chen2}$ where we have considered tunneling to/from a Weyl point nanostructure to discrete electron states. In the following, we will neglect the energy dependence of $\Gamma_{u,v}$ which is a common assumption for the tunneling at energies close to the Fermi energy.

With this, we can straightforwardly evaluate the rates of all relevant processes. Those include transitions between $|g\rangle$ and doublet states, $|e\rangle$ and doublet states, each transition can proceed with addition of spin $\sigma$ and either electron or hole to the lead. Let us consider a transition from $|x\rangle$ to $|-\sigma\rangle$ with addition of an electron with spin $\sigma$ to the lead. This should involve $\Gamma_u$. The energy of the resulting electron is $E = E_e - E_{-\sigma}$, and the probability to find an empty state for this transition is defined by the filling factor in the lead at the energy $E$ and with spin direction $\sigma$. Therefore, 
\begin{equation}
\Gamma_{x\to -\sigma, e} = \Gamma_u \left(1 -f_{\sigma}(E_x - E_{-\sigma})\right)
\end{equation}  
The other rates are obtained by similar consideration. Let us list them all (here for brevity $\bar{f} \equiv (1-f)$) :
\begin{eqnarray}
\Gamma_{x\to \sigma,e} &=& \Gamma_u \bar{f}_{-\sigma} (E_x - E_{\sigma})\\
\Gamma_{x \to \sigma,h} &=& \Gamma_v f_{\sigma}(E_\sigma - E_x)\\
\Gamma_{\sigma\to x,e} &=&  \Gamma_v \bar{f}_{\sigma} (E_{\sigma}-E_x)\\
\Gamma_{\sigma \to x,h} &=& \Gamma_u f_{-\sigma}(E_x - E_\sigma)\\
\Gamma_{g\to \sigma,e} &=&  \Gamma_v \bar{f}_{-\sigma} (E_g - E_{\sigma})\\
\Gamma_{g \to \sigma,h} &=&  \Gamma_u f_{\sigma}(E_\sigma - E_g)\\
\Gamma_{\sigma\to g,e} &=&  \Gamma_u \bar{f}_{\sigma} (E_{\sigma}-E_g)\\
\Gamma_{\sigma \to g,h} &=&  \Gamma_v f_{-\sigma}(E_g - E_\sigma)
\end{eqnarray}

We note that since $E_g = - E_x$ and $E_\sigma=-E_{-\sigma}$, this is the manifestation of the absence of electron-electron interactions in our model, 
\begin{equation}
\Gamma_{\sigma \to x} = \Gamma_{g \to -\sigma}; \; \Gamma_{x \to \sigma} = \Gamma_{-\sigma \to g}
\end{equation}
for both $e$ and $h$ processes separately. One can easily include more leads into the consideration: each rate will be a sum of contributions of the rates to each lead.

The rates will enter a standard master equation for the probabilities $p_g, p_x, p_\uparrow, p_\downarrow$. We will not write down the equation, since owing to the absence of the interactions, its solution is easily obtained in a very general situation and reads ($\bar{F}= 1-F$):
\begin{equation}
p_g = \bar{F}_u \bar{F}_d, \, p_\downarrow = F_d \bar{F}_u, \, p_\uparrow = F_u \bar{F}_d, \, p_x = F_u F_d, 
\end{equation}
where the effective "filling factors" $F_{d,u}$ are given by
\begin{eqnarray}
F_d = \Sigma^{-1}\sum_{j} \left(\Gamma^{(j)}_u f_\downarrow^{(j)}(\epsilon_d) + \Gamma^{(j)}_v\bar{f}_\uparrow(-\epsilon_d)\right); \\
F_u = \Sigma^{-1}\sum_{j} \left(\Gamma^{(j)}_u f_\uparrow^{(j)}(\epsilon_u) + \Gamma^{(j)}_v\bar{f}_\downarrow(-\epsilon_u)\right); \\
\epsilon_{u,d} = E_{\uparrow,\downarrow}-E_g; \; \Sigma \equiv \sum_j \left(\Gamma^{(j)}_u+\Gamma^{(j)}_v\right)
\end{eqnarray}%
$j$ being metallic lead index.

We mostly concentrate on the vanishing temperature case, $k_B T \ll B$. Then in the absence of spin accumulation the filling factor does not depend on spin and can be approximated $f(E) = \Theta(-E+eV)$, $V$ being the voltage applied to the lead. It is convenient to set the superconducting nanostructure at zero voltage.

\section{Single-lead transport}
\label{sec:sing}
Let us concentrate on a single lead setup and evaluate the current at various voltages applied to the lead. 
\begin{figure}
\begin{center}
\includegraphics[width=\columnwidth]{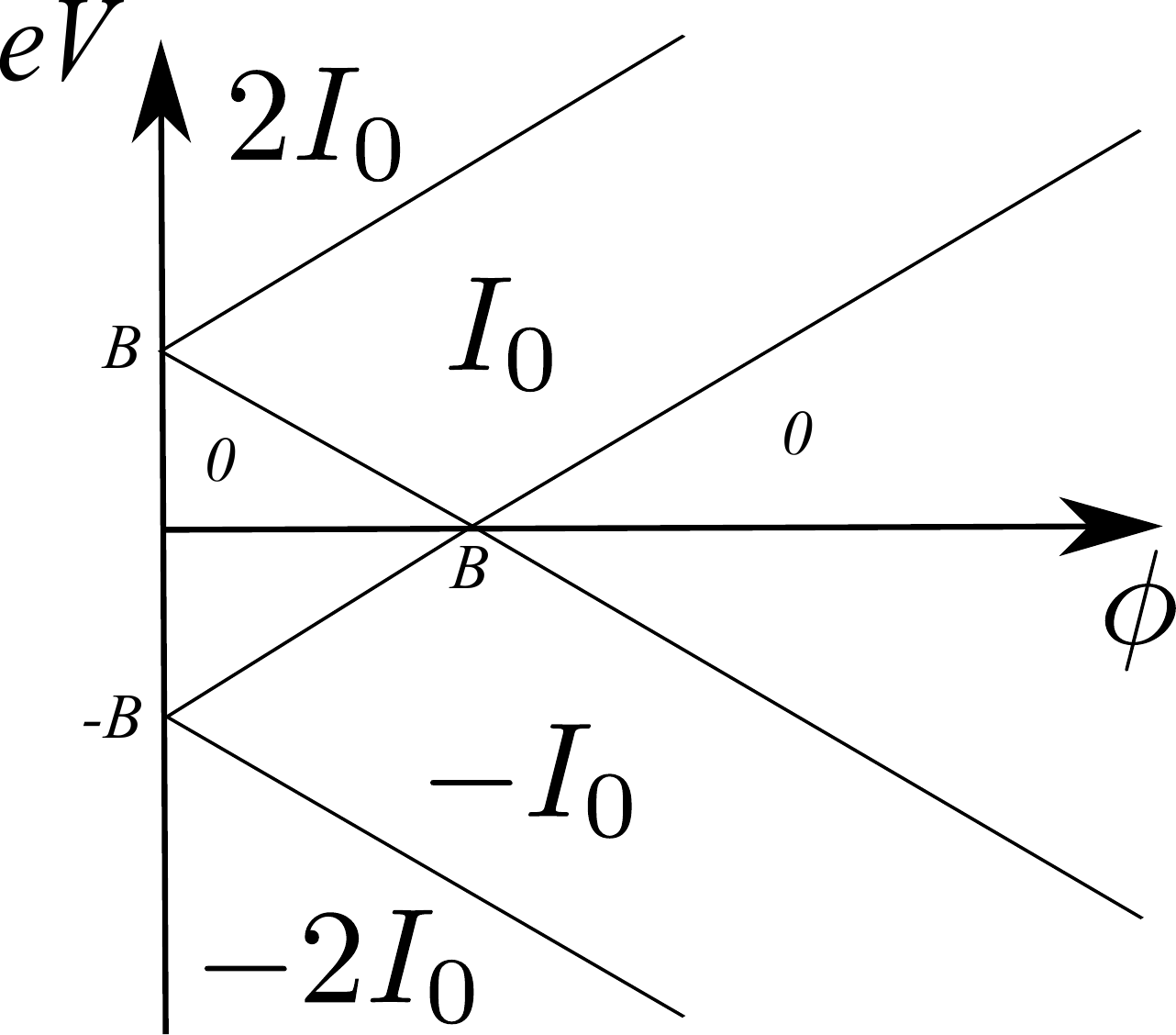}
\end{center}
\caption{\label{fig:onelead}Transport in the single-lead setup. The lines of thresholds at $eV=\pm|B-\phi|$ and $eV=\pm (B+\phi)$ define the domains with the electric current $I = \pm I_0, \pm 2 I_0$. There is no current at low voltage except $\phi \approx B$ and vanishing temperature. }
\end{figure}
To understand the relevant transport processes, let us fist assume vanishing temperature, $eV>0$ and $\phi>B$, that is, the singlet ground state. No current will flow until the voltage exceeds the threshold required to put a quasiparticle with spin down to the nanostructure, $eV > \epsilon_d = \phi-B$. At slightly higher voltage, the states at energy $e_d$ are filled in the lead and an electron at this energy tunnels to the nanostructure adding a quasiparticle. The rate of this process is $\Gamma_u$. The second quasiparticle can not be added yet since it requires higher energy. The state of the nanostructure only changes when an electron at energy $-\epsilon_d$ enters annihilating the quasiparticle. This process occurs with the rate $\Gamma_v$. Then the transport cycle repeats itself. We thus have two electrons transferred per cycle of the average duration $\Gamma^{-1}_{u}+ \Gamma^{-1}_{v}$, so that the current in this regime is given by
\begin{equation}
\label{eq:I0}
I = e \frac{2 \Gamma_u \Gamma_v}{\Gamma_u + \Gamma_v} \equiv I_0.
\end{equation}
If we start with the magnet ground state, $\phi <B$, the threshold voltage for the same transport regime is determined by opening the pair annihilation process, $eV > - \epsilon_d$. Both thresholds are combined in one by relation $eV>|\phi-B|$. (Fig. \ref{fig:onelead})
Upon further increase of voltage, we achieve another threshold $eV > \epsilon_d = B+\phi$ where electrons coming to the leads can add a quasiparticle with spin up, either to ground or spin-down state. Owing to the absence of interaction, this opens up another equivalent and independent transport channel, and the current doubles in this regime (Fig. \ref{fig:onelead}):
\begin{equation}
I= 2 I_0
\end{equation}

An interesting feature in this regime is a singular dependence of the current at the Weyl point $\phi=0$. Indeed, the current is a function of $\vec{n}$ (see Eq. \ref{eq:angular}),
\begin{equation}
I = e \frac{4 \Gamma_u \Gamma_v}{\Gamma_u +\Gamma_v} = e \Gamma_0 + e\frac{(\vec{\Gamma}_1 \cdot \vec{n})^2}{\Gamma_0}
\end{equation} 
At $\phi=0$, an infinitesimally small change of $\vec{\phi}$ leads to a finite change of the current. Remarkably, such divergent admittance response persist at finite voltages. In reality, the singularity is probably smoothed at $\phi \simeq \Gamma$, elaboration on this being beyond the approach of this article. Nevertheless, this anomalously big response can be used for a simple and reliable identification of the Weyl point position in a realistic experiment.

At negative $eV$, all the processes are accompanied by electrons leaving the nanostructure rather than entering it. This reverses the sign of the current upon reverting the voltage.

There is a relatively simple expression for the current beyond the vanishing temperature limit,
\begin{align}
I/I_0 &= (f_F(\epsilon_d -eV) - f_F(\epsilon_d + eV)) \notag
\\
&\qquad+ (f_F(\epsilon_u -eV) - f_F(\epsilon_u + eV))
\end{align}
where $f_F(\epsilon) \equiv (1+\exp(\epsilon/k_BT))^{-1}$ is the Fermi distribution function, two terms correspond to quasiparticle transfer with down or up spin.
At finite but small temperature $k_B T \ll B$  the zero-voltage conductance exhibits a resonant peak in the vicinity of $B=\phi$, that is, at $\epsilon_d \simeq k_B T \ll B$
\begin{equation}
\label{eq:resonance}
dI/dV = \frac{e I_0 }{2 k_B T} \frac{1}{{\rm cosh}^2(\epsilon_d/2 k_BT)}  
\end{equation}
This can be used for identification of the transition to the magnetic state.

There is no dc spin current in the sigle-lead setup owing to a simple fact: the current to the singlet superconductor bears no spin. In the next Sections, we show how this can be circumvented. 
   
\section{Spin on demand and a.c. spin current}
\label{sec:spino}
Let us understand that despite the fact that the dc spin current is absent for the single-lead setup, the spin injection is easy to organize.
Suppose we want a spin on demand: single spin injected to the lead in a time window around a time moment $t_0$. We can do so by changing the superconducting phases, that is, $\phi$. Before $t_0$, we keep $\phi>B$ so the state is the ground singlet. At $t=t_0$, we switch $\phi$ to a value $<B$ making down state energetically favorable. Within a time interval $\simeq \Gamma_0$ a spin will be injected to the lead, either as an electron or hole excitation.
To inject a spin of opposite sign, we keep $\phi<B$ before $t_0$ and change it to the value $B>0$. 

An obvious drawback of this scheme is that we cannot inject the spin of the same sign twice: we would need to evacuate the quasiparticle somewhere. In the single-lead setup, it would have to go to the same lead injecting the opposite spin. This drawback becomes an advantage if the goal is to produce an a.c. spin current $J$. 

\begin{figure}
\begin{center}
\includegraphics[width=\columnwidth]{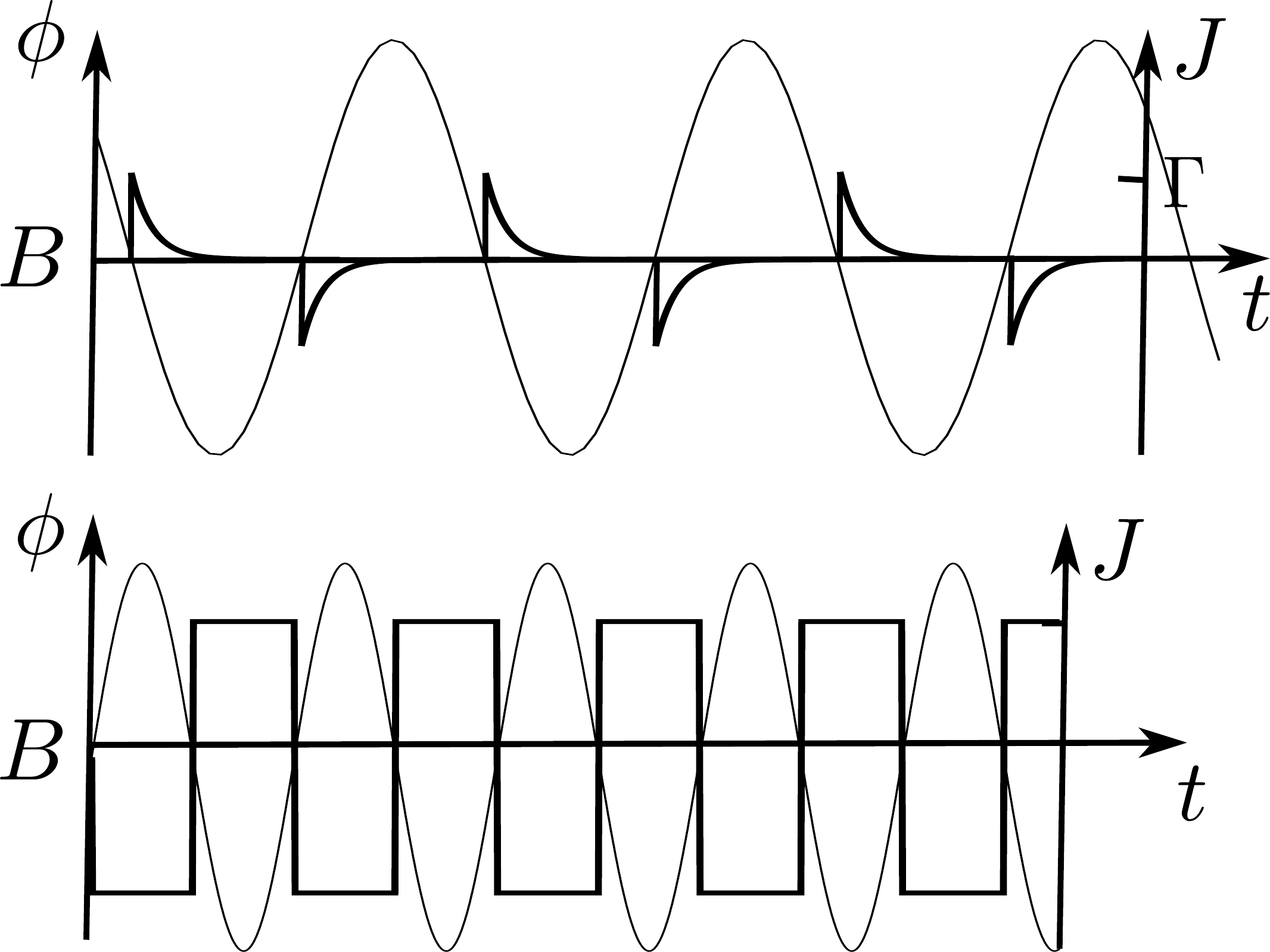}
\end{center}
\caption{\label{fig:acspin}Time-averaged a.c. spin current (thick curve) from the lead produced by a periodic modulation of the distance $\phi$ (thin curve) from the Weyl point. Upper plot: low frequencies $\Omega\ll \Gamma$, spin transfers at the time scale $\simeq \Gamma$ upon crossing the boundary of magnetic state region. Lower plot: high frequencies $\Omega\gg\Gamma$, equal populations of spin-down and ground singlet state.  }
\end{figure}

Suppose we cycle $\phi$ in the following way:
\begin{equation}
\phi(t) = B + \tilde{\phi} \sin(\Omega t)
\end{equation}
In the limit of low frequencies $\Omega < \Gamma_0$ , we have alternating single-spin injections at time moments $t_n = 2\pi n /\Omega$ (Fig. \ref{fig:acspin}). In the opposite limit of high frequencies $\Omega>\Gamma_0$, the down and ground singlet state are equally populated, the spin transfers are stochastic with the time-averaged spin current being given by 
\begin{equation}
J(t) = - {\rm sgn}\sin(\Omega t)\ \Gamma/2
\end{equation} 
Even in the limit of high frequencies, the amplitude of this a.c. spin current is comparable with d.c. spin currents we will evaluate later.

\section{Two-lead transport}
\label{sec:twol}
Let us start our discussion of the transport in the two-lead setup with a simple but perhaps the most interesting example. Let us organize an absolute spin-valve, that is, the transport involving only electrons of a single spin direction.
\begin{figure}
\begin{center}
\includegraphics[width=\columnwidth]{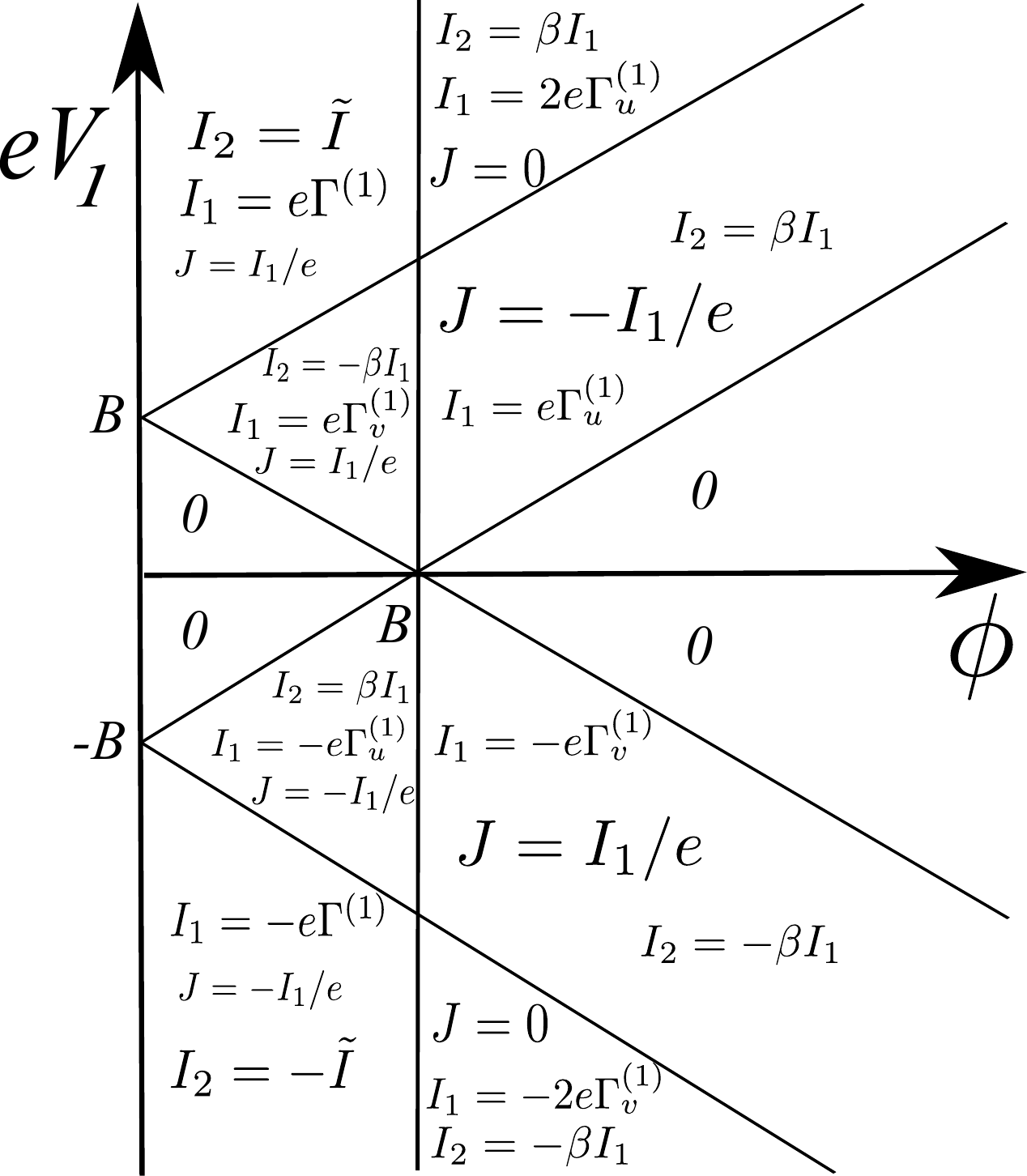}
\end{center}
\caption{\label{fig:abs}Two-lead setup. The absolute spin-valve regime can be realized at small $|eV_2|<|\epsilon_d|$ and $\Gamma^{(1)}_{u,v} \ll \Gamma^{(2)}_{u,v}$. Except the regions with zero spin current $J=0$, only electrons with either spin-down or spin-up are transported in the first lead. In all regions, $J=J_1=-J_2$. The current in the second lead is not completely polarized, $I_2 =\pm \beta I_1$ with $\beta \equiv  (\Gamma^{(2)}_{v}-\Gamma^{(2)}_{u})/(\Gamma^{(2)}_{v}+\Gamma^{(2)}_{u})$, or $I_2 =\pm \tilde{I}$, $\tilde{I} = e(\Gamma^{(1)}_{v}-\Gamma^{(1)}_{u})\beta$. }
\end{figure}
The tunneling to/from the two leads is characterized by the rates $\Gamma^{(1)}_{u,v}$, $\Gamma^{(2)}_{u,v}$. We assume vanishing temperature and $\phi>B$. We also set $V_2 =0$ and increase the voltage of the first lead. Nothing happens till $eV_1 <\epsilon_d$: the nanostructure remains in the ground singlet state. Upon crossing this threshold, spin-down electrons from the first lead can create a quasiparticle in the nanostructure. The quasiparticle can go either to the first or two the second lead. Let us assume $\Gamma^{(1)}_{u,v} \ll \Gamma^{(2)}_{u,v}$. In this case, the created quasiparticle will go to the second lead almost instantly bringing the nanostructure back to the ground singlet state. Therefore the transport in the first lead will involve only spin-down electrons, $I_1 = e \Gamma^{(1)}_u$, $J_1 = -I_1/e$. The absolute spin-valve is realized.

The spin current in the second lead is exactly opposite, $J_2 = - J_1$. As to the electric current, the quasiparticle decaying to this lead can create both electron and hole excitations. So that the current in the second lead is smaller in magnitude than $I_2$ and can be of either sign depending on the direction near the Weyl point,
\begin{equation}
I_2 = I_1 (\Gamma^{(2)}_{v}-\Gamma^{(2)}_{u})/(\Gamma^{(2)}_{v}+\Gamma^{(2)}_{u})=- 2 I_1 (\vec{\Gamma}^{(2)} \cdot \vec{n})/\Gamma^{(2)}.
\end{equation}

It is easy to revert the direction of the spin current. If $\phi <B$, the ground state is spin-down doublet and the transport in the first lead involves the spin-up electrons only, $I_1 = e \Gamma^{(1)}_v$, $J_1 = I_1/e$. The currents in the second lead follow $I_1, J_1$ as in the previous case. 

If we rise $eV_1$ above the second threshold, $eV_1>\epsilon_u$, at $\phi>B$ the quasiparticles with both spins can be created in the nanostructure, eventually, with equal probability. This quenches the spin current in this regime, while the electric current $I_1 = 2 e \Gamma^{(1)}_u$ is doubled. If $\phi <B$, the crossing of the second threshold does not change the absolute spin valve regime since the transitions from the spin-down state to either ground or excited singlet are both accompanied by the same spin change. The current increases to $I_1 = e \Gamma^{(1)}$  

Reverting $V_1$ changes the sign and magnitude of $I_{1}$ while $J$ follows the magnitude but remains of the same sign. The results for the absolute spin valve regime are summarized in Fig \ref{fig:abs}. At vanishing temperature, the transport is the same through the range $-|\epsilon_{d}| <eV_2< |\epsilon_d|$.

General picture of the transport in the two-lead setup beyond the assumption $\Gamma^{(1)} \ll \Gamma^{(2)}$ is more complex. The polarization of the transport electrons is not absolute. For instance, in the region defined by $-|\epsilon_{d}| <eV_2< |\epsilon_d|$, $\phi > B$, $\phi-B <eV_1<\phi+B$ the currents read:
\begin{eqnarray}
J&=&J_1=-J_2 = - \Gamma^{(1)}_u\frac{\Gamma^{(2)}}{ \Gamma^{(2)}+\Gamma^{(1)}} \\
I_1&=&e\Gamma^{(1)}_u\frac{\Gamma^{(2)}+ 2 \Gamma^{(1)}_v }{ \Gamma^{(2)}+\Gamma^{(1)}}\\
I_2&=& -e\Gamma^{(1)}_u\frac{\Gamma^{(2)}_u -\Gamma^{(1)}_v }{ \Gamma^{(2)}+\Gamma^{(1)}}
\end{eqnarray}
The polarization in the second lead is thus
\begin{equation}
|\frac{eJ}{I_1}| =  \frac{1}{1+2 \Gamma^{(1)}_v/\Gamma^{(2)}} <1
\end{equation}
In all voltage regions and arbitrary temperatures the currents are obtained from the general formulas 
\begingroup
\allowdisplaybreaks
\begin{align}
(\Gamma^{(2)}+ &\Gamma^{(1)})J =\notag
\\
& 
\Gamma^{(1)} \Gamma^{(2)}_u (f_F(\epsilon_d -eV_2) -f_F(\epsilon_u - eV_2)) \notag
 \\
&- \Gamma^{(2)} \Gamma^{(1)}_u (f_F(\epsilon_d -eV_1) -f_F(\epsilon_u - eV_1)) \notag
\\
&+ \Gamma^{(1)} \Gamma^{(2)}_v (f_F(\epsilon_d +eV_2) -f_F(\epsilon_u + eV_2) \notag 
\\
& - \Gamma^{(2)} \Gamma^{(1)}_v (f_F(\epsilon_d +eV_1) -f_F(\epsilon_u + eV_1)) 
 \\
(\Gamma^{(2)}+ &\Gamma^{(1)})I_1/e = \notag
\\ 
&2\Gamma^{(1)}_u \Gamma^{(1)}_v(f_F(\epsilon_d-eV_1) - f_F(\epsilon_d+eV_1) \notag
\\
&\qquad + f_F(\epsilon_u-eV_1) - f_F(\epsilon_u+eV_1)) \notag
\\
&+\Gamma^{(1)}_u \Gamma^{(2)}_u (f_F(\epsilon_d-eV_2) - f_F(\epsilon_d-eV_1) \notag
\\
&\qquad + f_F(\epsilon_u-eV_2) - f_F(\epsilon_u-eV_1)) \notag
\\
&+\Gamma^{(1)}_u \Gamma^{(2)}_v (f_F(\epsilon_d+eV_2) - f_F(\epsilon_d-eV_1) \notag
\\
&\qquad + f_F(\epsilon_u+eV_2) - f_F(\epsilon_u-eV_1)) \notag
\\
& +\Gamma^{(1)}_v \Gamma^{(2)}_u (f_F(\epsilon_d-eV_2) - f_F(\epsilon_d+eV_1) \notag
\\
&\qquad + f_F(\epsilon_u-eV_2) - f_F(\epsilon_u+eV_1)) \notag
\\
&+ \Gamma^{(1)}_v \Gamma^{(2)}_v (f_F(\epsilon_d+eV_2) - f_F(\epsilon_d+eV_1) \notag
\\
&\qquad + f_F(\epsilon_u+eV_2) - f_F(\epsilon_u+eV_1)) 
\end{align}
\endgroup

\section{Detection of spin accumulation}
\label{sec:dete}
So far we have considered equilibrium electron distribution in the normal leads. It is plausible to arrange a distribution that is not in equilibrium with respect to spin. \cite{SpintronicsReview,BauerTwo}
For instance, there may be another contact with this lead, that injects spin utilizing the properties of a traditional normal metal - ferromagnet interface. Owing to the approximate spin conservation, the distributions of the spins of two different directions can be regarded as independent and may differ in chemical potentials. This difference $2 P$ characterizes spin accumulation in energy units. If we assume thermalization of the distributions, the filling factors read
\begin{equation}
f_{\downarrow,\uparrow}(\epsilon) = f_F(\epsilon \pm P).
\end{equation}
If the axis of the resulting spin accumulation $\vec{P}$ is not in the direction of $\vec{B}$, the effective filling factors for two spin directions read
\begin{equation}
\label{eq:spinaccumulation}
\left[ \begin{array}{cc}
f_{\downarrow}(\epsilon) \\
f_{\uparrow}(\epsilon) \end{array}\right]
 =  \left[ \begin{array}{cc}
\cos^2\frac{\theta}{2} & \sin^2\frac{\theta}{2}\\
\sin^2\frac{\theta}{2} &\cos^2\frac{\theta}{2} 
 \end{array}\right]  \left[ \begin{array}{cc}
 f_F(\epsilon + P) \\ f_F(\epsilon - P)
 \end{array}\right]
\end{equation}
$\theta$ being the angle between $\vec{P}$ and $\vec{B}$.

A common spintronic effect is an electric current response on spin accumulation at one side of a contact.\cite{SpintronicsReview} This response may be present even without a voltage difference applied to the contact owing to spin dependence of the transmission coefficients. \cite{BauerFirst} It provides a convenient way to detect and measure the spin accumulation.

Let us start with the single-lead setup. In this case, the spin accumulation gives no current at zero voltage despite the difference in transport of spin-down and spin-up electrons. The reason for this is a rather fine symmetry of the distribution given by Eq. \ref{eq:spinaccumulation}: $f_\sigma(\epsilon) = \bar{f}_{-\sigma}(-\epsilon)$. This guarantees equal amount of electron emission and absorption by the superconducting nanostructure and thus zero net current.
\begin{figure}
\begin{center}
\includegraphics[width=\columnwidth]{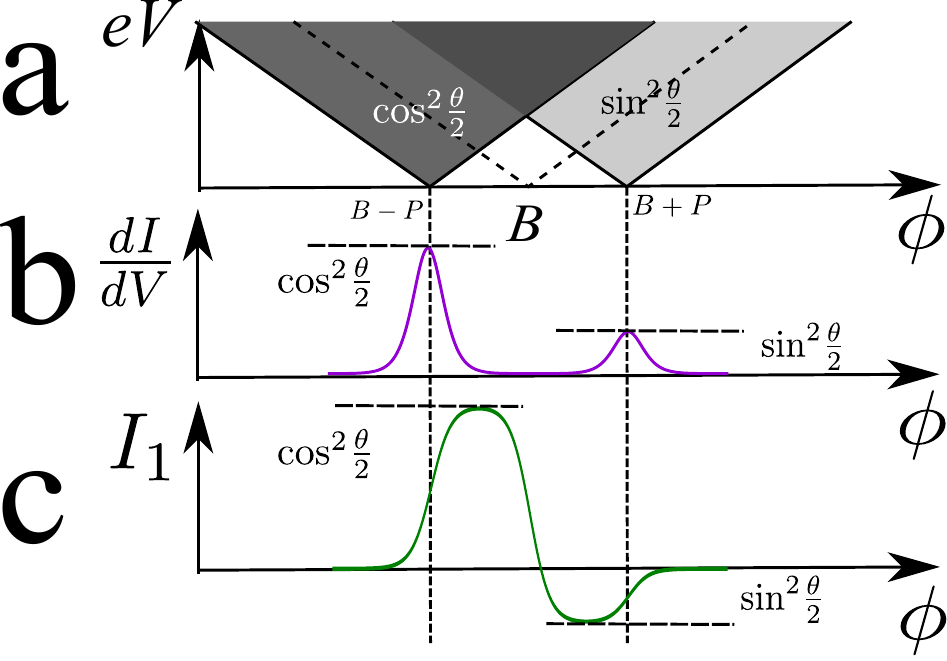}
\end{center}
\caption{\label{fig:detection}The detection of spin accumulation. a. The single-lead setup. The domain of the current with $I=I_0$ (see Fig.\ref{fig:onelead}) is spit into two corresponding to majority and minority spin accumulated, those are shifted  in $\phi$ by $\pm P$. The currents in the resulting domains are  $\cos^2\frac{\theta}{2} I_0$, $\sin^2\frac{\theta}{2} I_0$. b. The detection in the single-lead setup requires voltage. Low-voltage conductance corresponding to a. gives two peaks that are well-separated provided $k_B T \ll P$. In the plot, $k_B T =0.1 P$. c. In the two-lead setup, the spin accumulation gives rise to a current in the absence of voltage. The current in the first lead exhibits two plateaux $\simeq \cos^2\frac{\theta}{2}, -\sin^2\frac{\theta}{2}$. Here, $k_B T =0.1 P$. }
\end{figure}
The spin accumulation in this setup is however detected in the presence of voltage. At vanishing temperature, each boundary between the regions of different current is split by the spin accumulation. Two resulting boundaries correspond to thresholds for the transport of minority/majority spin and are shifted by $\pm P$ in $eV$, as shown in Fig. \ref{fig:detection} a. At finite temperature and small voltage, spin accumulation is manifested in splitting and $\pm P$ shifts of the conductance peak. Two separate peaks are formed if the accumulation is not in the direction of $\vec{B}$, otherwise the conductance peak is shifted by $P$, 
\begin{equation}
\frac{dI}{dV} = \frac{eI_0}{k_B T} \Big(\frac{\cos^2\frac{\theta}{2}}{{\rm cosh}^2((\epsilon_d+P)/2 k_BT)}+ \frac{\sin^2\frac{\theta}{2}}{{\rm cosh}^2((\epsilon_d-P)/2 k_BT)}\Big),
\end{equation} 
c.f. Eq. \ref{eq:resonance}, see also Fig. \ref{fig:detection}b. 
More general expression for the current reads
\begin{align}
I/I_0&=\cos^2\frac{\theta}{2} A_+ + \sin^2\frac{\theta}{2} A_- \,; \\
A_{\pm}&= f(\epsilon_d \pm P - eV) - f(\epsilon_d \pm P + eV) \notag
\\
&\qquad + f(\epsilon_u \mp P - eV) - f(\epsilon_u \mp P + eV) \notag
\end{align}

Interestingly, in a two-lead setup the spin accumulation is detected as a current signal without the voltages applied. We assume the spin accumulation is in the first lead. The accumulation $P < B$ gives rise to the current response near $\phi = B$ (Fig.\ref{fig:detection}c.) in the window $|\phi-B|<2P$.
In this regime, we can disregard the contribution of the spin-up excitations. The current in the first lead reads
\begin{align}
I_1 &= \frac{e \Gamma^{(1)} \Gamma^{(2)}}{\Gamma^{(1)} +\Gamma^{(2)}}\Big[\cos^2\frac{\theta}{2}(f_F(e_d+P)-f_F(\epsilon_d)) \notag
\\
 &\qquad\qquad +\sin^2\frac{\theta}{2}(f_F(e_d-P)-f_F(\epsilon_d)\Big] 
\end{align}
and $I_2=-I_1$.
Finally, we notice that the current response on the spin accumulation also remains the limit of high temperatures $k_B T \gg B, P$ where it is small in comparison with $I_0$ and linear in $\vec{P}$:
\begin{equation}
I_1 = -\frac{e \phi \Gamma^{(1)} \Gamma^{(2)}}{8(\Gamma^{(1)} +\Gamma^{(2)})(k_BT)^3} \left( 2 \vec{P}\cdot \vec{B} + \vec{P}^2 \right)
\end{equation}

\section{Conclusion}
\label{sec:conc}
To conclude, we have investigated transport from the normal leads to a superconducting nanostructure housing a Weyl point. A minimum magnet state is realized in the vicinity of this point. Owing to this, the transport exhibit all fundamental spintronic effects: the magnetic state can be detected, spin-on-demand and a.c. spin currents can be arranged in single-lead setups, spin-polarized current can be produced in two-lead setups, this includes the absolute polarization, the spin accumulation in a lead can be detected by electric measurement. The experimental realization of the setup and the corresponding spintronic experiments are feasible. Such a minimum spintronic device will be a demonstration of the power of superconducting nanotechnology and is advantageous because of its sensitivity to small changes of superconducting phase differences and energy selectivity of the transport.  

\begin{acknowledgements}
We acknowledge useful discussions with Alexander Balatsky. This research was supported by the European
Research Council (ERC) under the European Union's
Horizon 2020 research and innovation programme (grant
agreement No. 694272).
\end{acknowledgements}

\bibliographystyle{apsrev4-1}
\bibliography{manuscript1}
\end{document}